\documentclass[conference]{IEEEtran}
\usepackage{amsmath}
\usepackage{cite}
\usepackage{graphicx}
\usepackage{epstopdf}
\usepackage{amsfonts,amsmath,amssymb}
\usepackage{graphicx}
\usepackage{url}
\usepackage{bm}
\usepackage{bbm}
\usepackage{subfigure}
\usepackage{stfloats}
\usepackage{citesort}
\usepackage{color}

\usepackage{algorithmic}
\usepackage{algorithm}

\DeclareMathOperator*{\argmax}{argmax}
\DeclareMathOperator*{\argmin}{argmin}

\newtheorem{theorem}{\textbf{Theorem}}

\newcommand{\Hnull}{\mathcal{H}_0}
\newcommand{\Halt}{\mathcal{H}_1}

\newcommand{\Honull}{\mathcal{{D}}_0}
\newcommand{\Hoalt}{\mathcal{{D}}_1}

\begin{document}

\title{Covert Communication with Finite Blocklength in AWGN Channels}

\author{\IEEEauthorblockN{Shihao Yan$^{\dag}$,  Biao He$^{\ddag}$, Yirui Cong$^{\dag}$, and Xiangyun~Zhou$^{\dag}$}
\IEEEauthorblockA{$^{\dag}$Research School of Engineering, The Australian National University, Canberra, ACT, Australia}
\IEEEauthorblockA{$^{\ddag}$Department of Electronic and Computer Engineering, Hong Kong University of Science and Technology, Hong Kong}
\IEEEauthorblockA{Emails: \{shihao.yan, yirui.cong, xiangyun.zhou\}@anu.edu.au, eebiaohe@ust.hk}}


\maketitle

\begin{abstract}
Covert communication is to achieve a reliable transmission from a transmitter to a receiver while guaranteeing an arbitrarily small probability of this transmission being detected  by a warden. In this work, we study the covert communication in AWGN channels with finite blocklength, in which the number of channel uses is finite. Specifically, we analytically prove that the entire block (all available channel uses)  should be utilized to maximize the effective throughput of the transmission subject to a predetermined covert requirement. This is a nontrivial result because more channel uses  results in more observations at the warden for detecting the transmission. We also determine the maximum allowable transmit power per channel use, which is shown to decrease as the blocklength increases. Despite the decrease in the maximum allowable transmit power per channel use, the maximum allowable total power over the entire block is proved to increase with the blocklength, which leads to the fact that the effective throughput increases with the blocklength.
\end{abstract}



\section{Introduction}

In future wireless networks, the demand for wireless data is growing at such a rate that requires 1000x today's capacity in the next five to ten years. Against this background, crucial concerns on the security and privacy of wireless communications are emerging since a large amount of confidential information (e.g., email/bank account information and password, credit card details) is transferred over wireless networks. In addition to the secrecy and integrity of the transmitted information, in some scenarios a user may wish to transmit messages over wireless networks without being detected. This is due to the fact that (for example) the exposure of this transmission may disclose the user's location information, which probably violates the privacy of the user. Therefore, covert communication is attracting an increasing amount of research interests recently (e.g., \cite{che2014reliable,bash2015hiding,he2017on}). In covert communication, a transmitter (Alice) intends to communicate with a legitimate receiver (Bob) without being detected by a warden (Willie), who is observing this communication.

In fact, covert communication was addressed by spread spectrum techniques in the early 20th century and a review on spread spectrum techniques can be found in \cite{simon1994spread}. However, the performance limit of covert communication has not been fully examined in the literature and recently attracts much research attention. Considering additive white Gaussian noise (AWGN) channels, a square root law has been derived in \cite{bash2013limits}, which states that Alice can transmit no more than $\mathcal{O}(\sqrt{n})$ bits in $n$ channel uses covertly and reliably to Bob.  Following \cite{bash2013limits}, the scaling constant of the amount of information with respect to the square root of $n$ was characterized for a broad class of discrete memoryless channels (DMCs) and AWGN channels in \cite{wang2016fundamental}. We note that this square root law requires a pre-shared secret to be established between Alice and Bob prior to Alice's transmission. This pre-shared secret is proved to be unnecessary for the square root law when the channel quality from Alice to Bob is higher than that from Alice to Willie, for binary symmetric channel (BSC) \cite{{che2013reliable}}, DMC \cite{bloch2016covert}, and AWGN channel \cite{bloch2016covert}.


In the square root law we have $\mathcal{O}(\sqrt{n})/n \rightarrow 0$ as $n \rightarrow \infty$, which states that the rate is asymptotically zero (i.e., the average number of bits that can be covertly and reliably transmitted per channel use asymptotically approaches zero). However, in some scenarios a positive rate has been proved to be achievable (e.g., \cite{che2013reliable,lee2014achieving,bash2014LPD,lee2015achieving,sobers2015covert,goeckel2016covert}). For example, it is proved that a positive rate can be obtained when Willie has uncertainty about the receiver noise variance in AWGN channels \cite{lee2015achieving,goeckel2016covert},  when Willie does not exactly know the receiver noise model in BSC channels \cite{che2013reliable}, or when Willie lacks knowledge of his channel characteristics in AWGN and block fading channels \cite{sobers2015covert,goeckel2016covert}. In addition to the noise or channel uncertainty, as proved in \cite{bash2014LPD} a positive rate can also be achieved when Willie has uncertainty on the time instant of the communication.

In the literature as seen in the aforementioned works, only \cite{lee2015achieving} mentioned the impact of finite samples (i.e., finite $n$) on the detection performance at Willie. It is numerically shown that with noise uncertainty at Willie there may exist an optimal number of samples that maximizes the communication rate subject to $\xi \geq 1 - \epsilon$, where $\xi$ is the sum of false positive and miss detection rates at Willie and $0<\epsilon \leq 1$ is an arbitrarily small number. Besides the detection performance at Willie, finite $n$ also has significant impact on the maximal achievable rate $R$ of the channel from Alice to Bob (i.e., the maximal achievable rate decreases as $n$ decreases for a fixed decoding error probability $\delta$) \cite{poly2010channel}, which has not been considered in the literature of covert communication (including \cite{lee2015achieving}). Therefore, the impact of finite $n$ on covert communication has not been well examined. This leaves a significant gap in our understanding of the performance limit of practical covert communication, since in practice the length of a codeword is always finite. For example, to achieve transmission efficiency (e.g., short delay) we may require the codeword to be short (e.g., in the order of $~100$ channel uses) for vehicle-to-vehicle communication or real-time video processing \cite{makki2015finite}.

\subsection{Our Contributions}

Considering AWGN channels, we study the impact of finite $n$ on both the maximal achievable rate at Bob and detection performance at Willie in covert communication. To this end, noting that the decoding error probability $\delta$ is not negligible when $n$ is finite, we first propose to adopt the effective throughput $\eta$ (i.e., $\eta = n R (1-\delta)$) subject to $\xi \geq 1 - \epsilon$, as the performance metric to evaluate covert communication. As can be seen from the definition of $\eta$, it explicitly captures the tradeoff among $n$, $R$, and $\delta$ for a given covert requirement.

We consider a maximum blocklength of $N$ channel uses, in which the covert information needs to be transmitted. Hence, the actual number of channel uses $n$ is constrained by $n \leq N$. Although a larger $n$ offers more observations to Willie for detecting the transmission, we analytically prove that the optimal value of $n$ that maximizes $\eta$ subject to the given covert requirement is $N$ (i.e., the entire block with all available channel uses). We also determine the maximum allowable transmit power per channel use (denoted by $P^{\ast}$) that achieves the maximum $\eta$. Our examination shows that $P^{\ast}$ decreases as $N$ increases, which is due to the fact that increasing $N$ forces Alice to allocate less power for each channel use to meet the covert requirement. Nevertheless, we show that the maximum allowable total transmit power (i.e., $NP^{\ast}$) increases as $N$ increases, which leads to the fact that the effective throughput of the communication from Alice to Bob increases as $N$ increases. The results in this paper, for the first time, provide important insights on the design of covert communication with a finite blocklength.

\emph{Notations:} Scalar variables are denoted by italic symbols. Vectors and matrices are denoted by lower-case and upper-case boldface symbols, respectively. Given a vector $\mathbf{x}$, $x[i]$ denotes the $i$-th element of $\mathbf{x}$. The expectation is denoted by $\mathbb{E}\{\cdot\}$ and $\mathcal{CN}(0,\sigma^2)$ denotes the circularly-symmetric complex normal distribution with zero mean and variance $\sigma^2$.

\section{System Model}\label{system_model}

\subsection{Channel Model}

\begin{figure}[!t]
    \begin{center}
        \includegraphics[width=0.8\columnwidth]{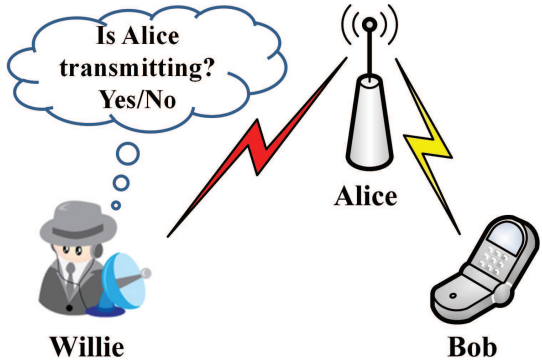}
        \caption{Illustration of the system model of interest for covert communication.}
        \label{fig:system}
    \end{center}
\end{figure}

The system model of interest for covert communication is illustrated in Fig.~\ref{fig:system}, where each of Alice, Bob, and Willie is equipped with a single antenna. We assume the channel from Alice to Bob and the channel from Alice to Willie are only subject to AWGN. In the covert communication, Alice transmits $n$ complex-valued symbols $x[i]$ ($i = 1, 2, \dots, n$) in each codeword to Bob, while Willie is passively collecting $n$ observations on Alice's transmission to detect her presence (i.e., whether Alice is transmitting). In this work, we consider that the length of a codeword is constrained by a maximum blocklength denoted by $N$. Thus, we have $n \leq N$ as a constraint on $n$.
We denote the AWGN at Bob and Willie as $r_b[i]$ and $r_w[i]$, respectively, where $r_b[i] \sim \mathcal{CN}(0,\sigma_b^2)$, $r_w[i] \sim \mathcal{CN}(0,\sigma_w^2)$, $\sigma_b^2$ and $\sigma_w^2$ are the noise variances at Bob and Willie, respectively. In addition, we assume that $x[i]$, $r_b[i]$, and $r_w[i]$ are mutually independent. We denote the transmit power of Alice as $P$ (i.e., $\mathbb{E}\{|x[i]|^2\} = P$). Furthermore, we assume that Alice adopts Gaussian signaling, i.e., $x[i] \sim \mathcal{CN}(0,P)$.



\subsection{Channel Coding Rate for Finite Blocklength}

The received signal at Bob for each signal symbol is given by
\begin{align}\label{xn}
y_b[i] = x[i] + r_b[i].
\end{align}
As pointed out by \cite{poly2010channel}, the decoding error probability at Bob is not negligible when $n$ is finite. As such, for a given decoding error probability $\delta$ the channel coding rate of the channel from Alice to Bob can be approximated by \cite{poly2010channel,ozcan2013throughput}
\begin{align}\label{rate_finite}
R \approx \log_2(1+\gamma_b) - \sqrt{\frac{\gamma_b(\gamma_b + 2)}{n(\gamma_b+1)^2}} \frac{{Q}^{-1}(\delta)}{\ln(2)} + \frac{\log_2(n)}{2n},
\end{align}
where $\gamma_b = {P}/{\sigma_b^2}$ is the signal-to-noise ratio (SNR) at Bob, and ${Q}^{-1}(\cdot)$ is the inverse Q-function. Equivalently, for a given channel coding rate $R$, the decoding error probability at Bob is given by
\begin{align}\label{delta_finite}
\delta = Q\left(\frac{\sqrt{n} (1+\gamma_b)\left(\ln(1+\gamma_b) + \frac{1}{2}\ln (n) - R \ln 2\right)}{\sqrt{\gamma_b(\gamma_b+2)}}\right).
\end{align}

\subsection{Binary Hypothesis Testing at Willie}

In order to detect Alice's presence, Willie is to distinguish the following two hypotheses
\begin{eqnarray}\label{hypothses}
 \left\{ \begin{aligned}\label{ncon}
        \ & \Hnull:~y_w[i] = r_w[i] \;\;\\
        \ & \Halt:~y_w[i] = x[i] + r_w[i],
         \end{aligned} \right.
\end{eqnarray}
where $\Hnull$ denotes the null hypothesis where Alice is not transmitting, $\Halt$ denotes the alternative hypothesis where Alice is transmitting, and $y_w[i]$ is the received signal at Willie. Following the assumptions detailed in Section II-A, we have the likelihood functions of $y_w[i]$ under $\Hnull$ and $\Halt$ as $f(y_w[i]|\Hnull) = \mathcal{CN}(0, \sigma_w^2)$ and $f(y_w[i]|\Halt) = \mathcal{CN}(0, P + \sigma_w^2)$, respectively. In the cover communication, the ultimate goal of Willie is to minimize the total error rate, which is given by
\begin{align}
\xi = P_{F} + P_{M},
\end{align}
where $P_{F} \triangleq \Pr(\Hoalt|\Hnull)$ is the false positive rate, $P_{M} \triangleq \Pr(\Honull|\Halt)$ is the miss detection rate, $\Hoalt$ and $\Honull$ are the binary decisions that infer whether Alice is present or not, respectively. We assume that Willie knows both $P$ and $\sigma_w^2$ exactly, and thus the optimal test that minimizes $\xi$ is the likelihood ratio test with $\lambda = 1$ as the threshold\footnote{We note that $\lambda = 1$ is due to the unknown or equal \emph{a priori} probabilities, i.e., $P_0$ and $P_1$ are unknown or equal, where $P_0$ is the \emph{a priori} probability that $\Hnull$ is true, $P_1$ is the \emph{a priori} probability that $\Halt$ is true, and $P_0 + P_1 = 1$. If both $P_0$ and $P_1$ are known, the total error rate is reformulated as $\xi = P_0 P_{F} + P_1 P_{M}$ and the optimal test that minimizes this reformulated $\xi$ is the likelihood ratio test with $\lambda = P_1/P_0$. We also note that the assumption of equal \emph{a priori} probabilities is commonly adopted in the literature of covert communication (e.g., \cite{bash2013limits,bash2014LPD,lee2015achieving}).}, which is given by
\begin{equation}\label{LRT}
\frac{\mathbb{P}_1 \triangleq \prod_{i = 1}^n f\left(y_w[i]|\Halt\right)}{\mathbb{P}_0 \triangleq \prod_{i = 1}^n f\left(y_w[i]|\Hnull\right)} \begin{array}{c}
\overset{\Hoalt}{\geq} \\
\underset{\Honull}{<}
\end{array}%
1.
\end{equation}
After performing some algebraic manipulations, \eqref{LRT} can be reformulated as
\begin{equation}\label{rodiometer}
T \triangleq \frac{1}{n} \sum_{i = 1}^n |y_w[i]|^2 \begin{array}{c}
\overset{\Hoalt}{\geq} \\
\underset{\Honull}{<}
\end{array}%
\Gamma,
\end{equation}
where $T$ is the average power of each received symbol at Willie and $\Gamma$ is the threshold for  $T$, which is given by
\begin{align}
\Gamma = \frac{(P + \sigma_w^2) \sigma_w^2}{P} \ln \left(\frac{P + \sigma_w^2}{\sigma_w^2}\right).
\end{align}
As per \eqref{LRT} and \eqref{rodiometer}, we note that the radiometer is indeed the optimal detector when Willie knows the likelihood functions exactly (i.e., there are no nuisance parameters embedded in the likelihood functions). Following \eqref{rodiometer} and noting that $T$ is a chi-squared random variable with $2n$ degrees of freedom, the false positive rate and miss detection rate are given by \cite{lee2014achieving,lee2015achieving}
\begin{align}
P_{F} &= \Pr(T > \Gamma|\Hnull) = 1- \frac{\gamma\left(n, \frac{n \Gamma}{\sigma_w^2}\right)}{\Gamma(n)} ,\label{false_positive_rate}\\
P_{M} &= \Pr(T < \Gamma|\Halt) = \frac{\gamma\left(n, \frac{n \Gamma}{P+\sigma_w^2}\right)}{\Gamma(n)},\label{detection_rate}
\end{align}
where $\Gamma(n) = (n-1)!$ is the gamma function and  $\gamma(\cdot, \cdot)$ is the incomplete gamma function given by
\begin{align}
\gamma(n, x) = \int_0^x e^{-t} t^{n-1} dt.
\end{align}

With the radiometer as the optimal detector, following Pinsker's inequality, we have a lower bound on $\xi$, which is given by \cite{bash2013limits,lehmann2005testing,cover2002elements}
\begin{align}\label{lowerbound_xi}
\xi \geq 1 - \sqrt{\frac{1}{2}\mathcal{D}(\mathbb{P}_0 \| \mathbb{P}_1)},
\end{align}
where $\mathcal{D}(\mathbb{P}_0 \| \mathbb{P}_1)$ is the Kullback-Leibler (KL) divergence from  $\mathbb{P}_0$ to $\mathbb{P}_1$, which can be expressed as
\begin{align}\label{KL}
\mathcal{D}(\mathbb{P}_0 \| \mathbb{P}_1) = n \left[\ln \left(\frac{P+\sigma_w^2}{\sigma_w^2}\right) - \frac{P}{P+\sigma_w^2} \right].
\end{align}

\subsection{Covert Requirement}

Covert communication requires $\xi \geq 1 - \epsilon$ for some arbitrarily small $\epsilon$. As per \eqref{lowerbound_xi}, we can ensure $\mathcal{D}(\mathbb{P}_0 \| \mathbb{P}_1) \leq 2 \epsilon^2$ in order to guarantee $\xi \geq 1 - \epsilon$. We note that $\mathcal{D}(\mathbb{P}_0 \| \mathbb{P}_1) \leq 2 \epsilon^2$ is a more strict constraint relative to $\xi \geq 1 - \epsilon$ as per \eqref{lowerbound_xi}. From a conservative point of view and to avoid the complex expressions for $P_{F}$ and $P_{M}$, in this work we adopt $\mathcal{D}(\mathbb{P}_0 \| \mathbb{P}_1) \leq 2 \epsilon^2$ as the requirement for covert communication. Also, the value of $\epsilon$ is especially very small in order to provide good covertness. Thus, in this work we only consider $\epsilon \in (0,0.5]$ because $\epsilon >0.5$ means that Willie is allowed to achieve more than 50\% success detection rate.

\section{Covert Communication with A Finite Number of Channel Uses}

In this section, we first adopt the effective throughput to evaluate the performance of covert communication in AWGN channels with finite blocklength. Then, we determine the optimal $n$ and $P$ that maximize this effective throughput subject to the covert requirement.

\subsection{Effective Throughput}

The square root law states that Alice can transmit no more than $\mathcal{O}(\sqrt{n})$ bits in $n$ channel uses covertly and reliably to Bob. Such scaling-law results are obtained when  $n \rightarrow \infty$. As such, these square-law results cannot be applied in the covert communication with finite $n$. In this work, we focus on the amount of information that can be transmitted reliably from Alice to Bob for a given positive $\epsilon$. Noting that the decoding error probability of a channel with finite blocklength is not negligible, we adopt the effective throughput from Alice to Bob as the main performance metric for the covert communication with finite blocklength, while utilizing the covert requirement as the constraint. The effective throughput from Alice to Bob is defined as \cite{yan2015optimization,zhou2011rethinking}
\begin{align}
\eta = n R (1-\delta). \label{throughput}
\end{align}
We note that $\eta$ gives the average number of information bits that can be transmitted from Alice to Bob reliably (excluding information bits suffering from decoding errors) by utilizing a codeword with finite length $n$.



\subsection{Optimal Number of Channel Uses and Transmit Power}


The ultimate goal of our design in covert communication is to achieve the maximum effective throughput while guaranteeing the covert requirement. To this end, we first consider a fixed channel coding rate $R$ and focus on the design of the number of channel uses and the transmit power $P$, since the design of $n$ and $P$ affects both the effective throughput from Alice to Bob and the detection performance at Willie. As such, for a given $R$ the optimization problem in the covert communication of interest can be written as
\begin{align}
 &\argmax_{n, P} \eta,\\
&~~~~\text{s.t.} ~~\mathcal{D}(\mathbb{P}_0 \| \mathbb{P}_1) \leq 2 \epsilon^2, \label{constrain} \\
&~~~~~~~~~~ n \leq N.
\end{align}

\begin{theorem}\label{theorem2}
The optimal values of $n$ and $P$ that maximize the effective throughput $\eta$ subject to $\mathcal{D}(\mathbb{P}_0 \| \mathbb{P}_1) \leq 2 \epsilon^2$ and $n \leq N$ are derived as
\begin{align}
n^{\ast} &= N, \\
P^{\ast} &= (\sigma_w^2 + P^{\ast})\left[\ln\left(\frac{P^{\ast}}{\sigma_w^2}+ 1\right) - 2 \epsilon^2 N\right], \label{fix_P}
\end{align}
where $P^{\ast}$ is the solution to the fixed-point equation \eqref{fix_P}.
\end{theorem}
\begin{IEEEproof}
The detailed proof is provided in Appendix.
\end{IEEEproof}

Based on Theorem~\ref{theorem2}, we see that it is best for Alice to transmit over all available channel uses for covert communication, provided that the transmit power is optimized to maintain the same level of covertness despite that Willie has more observations when $n$ is larger. The same level of covertness is achieved by reducing the transmit power when $n$ becomes larger, which can be seen from \eqref{fix_P} that $P^{\ast}$ decreases with $N$. It is interesting to observe that both $n^{\ast}$ and $P^{\ast}$ are not functions of $R$. This demonstrates that the obtained $n^{\ast}$ and $P^{\ast}$ are globally optimal, regardless the value of the channel coding rate $R$. As such, the optimal value of $R$ that maximizes the effective throughput subject to the covert requirement can be obtained through
\begin{align}
R^{\ast} =  \argmin_{0 \leq R} N R \left[1-\delta(P^{\ast}, N, R)\right],
\end{align}
where $\delta(P^{\ast}, N, R)$ is obtained by substituting $P = P^{\ast}$ and $n^{\ast} = N$ into \eqref{delta_finite}. We note that $R^{\ast}$ can be also obtained through searching the optimal value of $\delta$ that maximizes $\eta$ for $n^{\ast} = N$ and $P = P^{\ast}$. We define $\delta^{\ast} = \delta(P^{\ast}, N, R^{\ast})$ and denote the maximum effective throughput as $\eta^{\ast}$, which is achieved by substituting $P^{\ast}$, $n^{\ast}$, $R^{\ast}$, and $\delta^{\ast}$ into \eqref{throughput}.

\section{Numerical Results}

In this section, we provide numerical results on the effective throughput subject to $\xi \geq 1 - \epsilon$ to verify our analysis on the covert communication with $\mathcal{D}(\mathbb{P}_0 \| \mathbb{P}_1) \leq 2 \epsilon^2$ as the constraint.

\begin{figure}[!t]
    \begin{center}
   {\includegraphics[width=3.4in, height=2.7in]{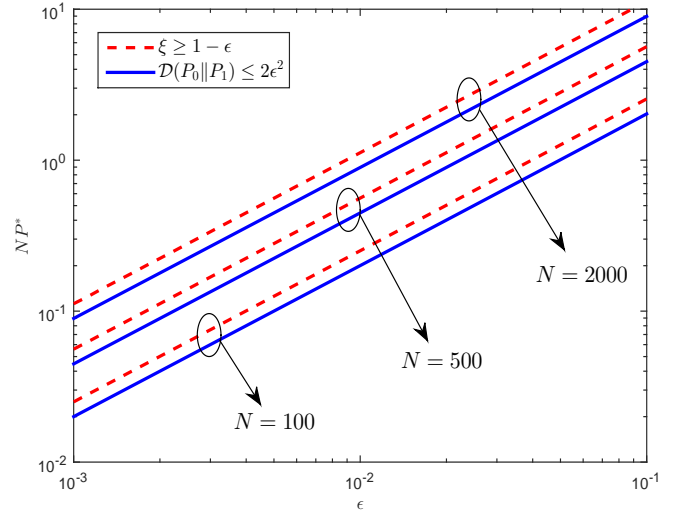}}
    \caption{Maximum allowable total transmit power $NP^{\ast}$ versus $\epsilon$ for different values of $N$, where $\sigma_b^2 =\sigma_w^2= 1$.}\label{fig:fig1}
    \end{center}
\end{figure}

In Fig.~\ref{fig:fig1} we plot the maximum allowable total transmit power $NP^{\ast}$ over the entire block versus $\epsilon$. In this figure and the following figures, the curves for $\xi \geq 1 - \epsilon$ are achieved by numerically evaluating the false positive and detection rates as per \eqref{false_positive_rate} and \eqref{detection_rate}, respectively. In this figure, we observe that the $NP^{\ast}$ with $\xi \geq 1 - \epsilon$ as the constraint is higher than that with $\mathcal{D}(\mathbb{P}_0 \| \mathbb{P}_1) \leq 2 \epsilon^2$ as the constraint. This is due to the fact that the equality in \eqref{lowerbound_xi} cannot be achieved in the considered system model, and hence $\mathcal{D}(\mathbb{P}_0 \| \mathbb{P}_1) \leq 2 \epsilon^2$ is a more strict constraint than $\xi \geq 1 - \epsilon$.
We also observe that $NP^{\ast}$ increases (hence the effective throughput increases) as $N$ increases, which can be explained by our Theorem~\ref{theorem2}.
Finally, we observe that $NP^{\ast}$ decreases (hence the effective throughput decreases) as $\epsilon$ decreases, which demonstrates the tradeoff between the covert requirement and the achievable effective throughput (e.g., a more strict covert requirement leads to a smaller effective throughput).

\begin{figure}[!t]
    \begin{center}
   {\includegraphics[width=3.4in, height=2.7in]{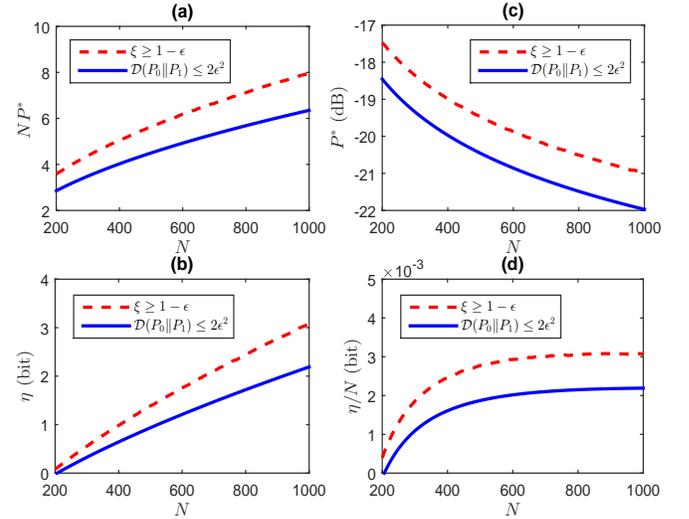}}
    \caption{$NP^{\ast}$, $\eta$, $P^{\ast}$, and $\eta/N$ versus $N$, where $\sigma_b^2 =\sigma_w^2= 1$, $\delta = 0.01$, and $\epsilon = 0.1$.}\label{fig:fig2}
    \end{center}
\end{figure}

In Fig.~\ref{fig:fig2}, we plot $NP^{\ast}$, $\eta$, $P^{\ast}$, and $\eta/N$ versus $N$ in different sub-figures, respectively. As expected, we first observe that $NP^{\ast}$ and $\eta$ monotonically increase as $N$ increases in Fig.~\ref{fig:fig2} (a) and Fig.~\ref{fig:fig2} (b), respectively.
Although $NP^{\ast}$ increases as shown in Fig.~\ref{fig:fig2} (a), it is interesting to observe that the maximum allowable transmit power $P^{\ast}$ monotonically decreases as $N$ increases in Fig.~\ref{fig:fig2} (c). This can be explained by \eqref{fix_P} in our Theorem~\ref{theorem2}. Intuitively, this is due to the fact that as the number of observations at Willie increases, Alice has to reduce her transmit power in order to meet the same detection performance at Willie. In Fig.~\ref{fig:fig2} (d), we observe that the effective throughput per channel use (i.e., $\eta/N$) monotonically increases as $N$ increases. This is due to the fact that the decrease in $\delta$ (i.e., the decoding error probability given in \eqref{delta_finite}) caused by increasing $N$ is more than the increase in $\delta$ caused by the reduction of $P^{\ast}$ as shown in Fig.~\ref{fig:fig2} (c). These aforementioned observations based on Fig.~\ref{fig:fig2} demonstrate that increasing $N$ not only helps Alice to allocate less transmit power to each channel use in order to maintain the same level of covertness, but also reduces the decoding error probability in the communication from Alice to Bob, which turns out to improve the effective throughput of the covert communication.

\begin{figure}[!t]
    \begin{center}
   {\includegraphics[width=3.4in, height=2.7in]{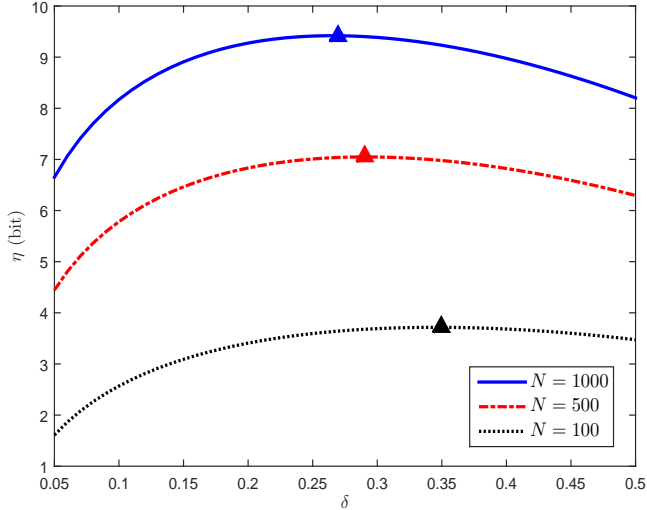}}
    \caption{Effective throughput $\eta$ versus the decoding error probability $\delta$ for different values of $N$, where $\sigma_b^2 =\sigma_w^2= 1$ and $\epsilon = 0.1$.}\label{fig:fig3}
    \end{center}
\end{figure}

\begin{figure}[!t]
    \begin{center}
   {\includegraphics[width=3.4in, height=2.7in]{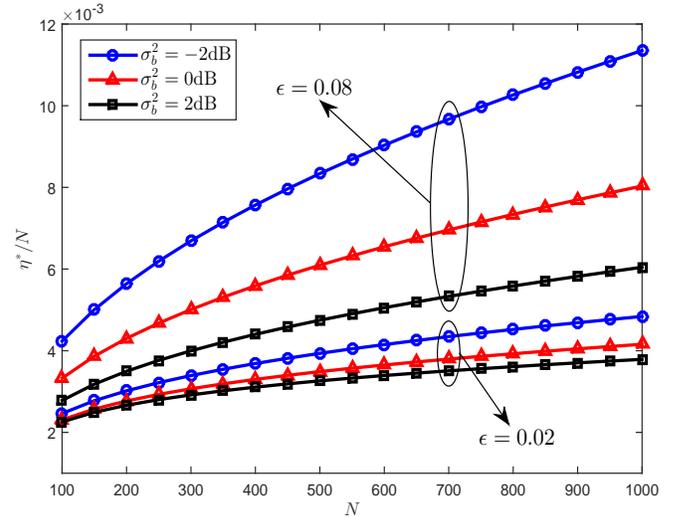}}
    \caption{Maximum effective throughput per channel use $\eta^{\ast}/N$ versus $N$ for different values of $\epsilon$ and $\sigma_b^2$, where $\sigma_w^2= 1$.}\label{fig:fig4}
    \end{center}
\end{figure}

In Fig.~\ref{fig:fig3}, we plot the effective throughput $\eta$ subject to $\xi \geq 1 - \epsilon$ versus the decoding error probability $\delta$. We first observe that the optimal value of $\delta$ that maximizes $\eta$ indeed exists, based on which we can determine the optimal $R$. We also observe that the optimal value of $\delta$ decreases as $N$ increases. As shown in Fig.~\ref{fig:fig2} (c), the maximum allowable  transmit power $P^{\ast}$ decreases as $N$ increases. As per \eqref{rate_finite}, the observation, that both $\delta^{\ast}$ and $P^{\ast}$ decreases as $N$ increases, indicates that the optimal channel coding rate $R^{\ast}$ decreases as $N$ increases. We also plot the maximum effective throughput per channel use (i.e., $\eta^{\ast}/N$) versus $N$ in Fig.~\ref{fig:fig4}. In this figure, we first observe that as $N$ increases $\eta^{\ast}/N$ increases, which is consistent with our observation found in Fig.~\ref{fig:fig2} (d). We also observe that as $\epsilon$ increases slightly (e.g., from $0.02$ to $0.08$) $\eta^{\ast}/N$ significantly increases. This demonstrates that the achievable effective throughput is very sensitive to the the covert requirement.

\section{Conclusion}\label{conclusion}

This work investigated the covert communication with finite blocklength (i.e., a finite number of channel uses $n \leq N$) over AWGN channels. We proved that the effective throughput of covert communication is maximized when all available channel uses are utilized, i.e., $n^{\ast} = N$. To guarantee the same level of covertness, the maximum allowable transmit power per channel use decreases as $N$ increases, while the maximum allowable  total transmit power over all channel uses increases as $N$ increases. In contrast, we found that both the effective throughput and the effective throughput per channel use increase as $N$ increases. This is due to the fact that increasing $N$ not only reduces the transmit power allocated to each channel use, but also decreases the decoding error probability of the communication from Alice to Bob.


\section*{Acknowledgments}

This work was supported by the Australian Research Council's Discovery Projects (DP150103905).

\section*{Appendix}

We present our proof of Theorem~\ref{theorem2} in the following 6 steps.

\textbf{Step 1:}
We note that $\eta$ and $\mathcal{D}(\mathbb{P}_0 \| \mathbb{P}_1)$ are both monotonically increasing functions of  $P$ and $n$. As such, we can conclude that the equality in the constraint \eqref{constrain} is always met in order to maximize $\eta$. Thus, we have $\mathcal{D}(\mathbb{P}_0 \| \mathbb{P}_1) = 2 \epsilon^2$ and following \eqref{KL} we have
\begin{align}\label{n_P}
n = \frac{2 \epsilon^2}{f(\gamma_w)},
\end{align}
where
\begin{align}\label{f_P}
f(\gamma_w) \triangleq \frac{\mathcal{D}(\mathbb{P}_0 \| \mathbb{P}_1)}{n}=  \ln \left({\gamma_w+1}\right) - \frac{\gamma_w}{\gamma_w+1},
\end{align}
and $\gamma_w = P/\sigma_w^2$ is the SNR at Willie.

\textbf{Step 2:}
We note $f(0) = 0$ and we derive the first derivative of $f(\gamma_w)$ with respect to $\gamma_w$ as
\begin{align}\label{f_P_d}
\frac{\partial f(\gamma_w)}{\partial \gamma_w} = \frac{\gamma_w}{(\gamma_w + 1)^2} \geq 0,
\end{align}
which leads to the fact that $f(\gamma_w)$ is a monotonically increasing function of $\gamma_w$.
With the constraint $\mathcal{D}(\mathbb{P}_0 \| \mathbb{P}_1) = 2 \epsilon^2$, $n$ is a monotonically decreasing function of $f(\gamma_w)$ as per \eqref{n_P}, which results in that $n$ is a monotonically decreasing function of $\gamma_w$ (thus of $P$).

\textbf{Step 3:}
Instead of directly proving $n^{\ast} = N$ for maximizing the effective throughput, we next prove that $n^{\ast} = N$ maximizes $n  \gamma_w$ (i.e., maximizes $nP$) under the constraint \eqref{n_P} in the remaining steps. This is due to the fact that  $n P$ is the total transmit power for the $n$ channel uses and the effective throughput increases as the total transmit power increases \cite{poly2010channel}.

\textbf{Step 4:}
We next prove that either $n = 1$ or $n = N$ maximizes $n \gamma_w$. To this end, in the following we first show that $n \gamma_w$ initially decreases and then increases with $n$. Following  \eqref{n_P} and \eqref{f_P}, we have
\begin{align}\label{np}
n \gamma_w = \frac{2 \epsilon^2}{g(\gamma_w)},
\end{align}
where $g(\gamma_w)$ is given by
\begin{align}
g(\gamma_w) = \frac{\ln(1+\gamma_w)}{\gamma_w} - \frac{1}{1+\gamma_w}.
\end{align}
We then derive the first derivative of $g(\gamma_w)$ with respect to $\gamma_w$ as
\begin{align}\label{g_derivative}
\frac{\partial g(\gamma_w)}{\partial \gamma_w} = \frac{h(\gamma_w)}{\gamma_w^2 (1+\gamma_w)^2},
\end{align}
where
\begin{align}
h(\gamma_w) = 2 \gamma_w^2 + \gamma_w - (1+\gamma_w)^2 \ln(1 + \gamma_w).
\end{align}
We note that there are \emph{only} two solutions to $h(\gamma_w) = 0$ for $\gamma_w \geq 0$, including $\gamma_w = 0$ and $\gamma_w = \gamma_w^{\dag}$.\footnote{We obtain $\gamma_w^{\dag} \approx 2.1626$ by numerically solving $h(\gamma_w) = 0$.}  We also note that as $\gamma_w \rightarrow \infty$ we have $h(\gamma_w) \rightarrow  -\infty$. Then, we can conclude that $h(\gamma_w)\geq 0$ for $\gamma_w < \gamma_w^{\dag}$ and $h(\gamma_w) \leq 0$ for $\gamma_w \geq \gamma_w^{\dag}$.
As such, noting $\gamma_w^2 (1+\gamma_w)^2 \geq 0$ and following \eqref{g_derivative}, we have $\partial g(\gamma_w)/\partial \gamma_w \geq 0$ for $\gamma_w < \gamma_w^{\dag}$ and $\partial g(\gamma_w)/\partial \gamma_w \leq 0$ for $\gamma_w \geq \gamma_w^{\dag}$. This indicates that $g(\gamma_w)$ initially increases and then decreases with $\gamma_w$. As per \eqref{np}, we know that $n \gamma_w$ monotonically decreases with $g(\gamma_w)$, which leads to the fact that $n \gamma_w$ first decreases and then increases as $\gamma_w$ increases (i.e., $n \gamma_w$ has one minimum value but no maximum value). We recall that $n$ is a monotonically decreasing function of $\gamma_w$ under the constraint \eqref{n_P}, which is proved following \eqref{f_P_d}. Therefore, we conclude that $n \gamma_w$ first decreases and then increases as $n$ increases, and thus the maximum value of $n \gamma_w$ is achieved either at $n = 1$ or $n = N$.

\textbf{Step 5:}
We next prove that $n = N$ (not $n = 1$) maximizes $n \gamma_w$. Substituting $\gamma_w^{\dag}$ into \eqref{n_P}, we have $n^{\dag} = 2 \epsilon^2/f(\gamma_w^{\dag})$.
For $0 < \epsilon < 0.4835$, we have $n^{\dag} < 1$ due to $f(\gamma_w^{\dag}) > 0.4675$. When  $n^{\dag} < 1$, $n \gamma_w$ increases with $n$ due to $n \geq 1$. As such, for $0 < \epsilon < 0.4835$ the optimal value of $n$ that maximizes $n \gamma_w$ is $N$ (i.e., $n^{\ast} = N$).
For $0.4835 \leq \epsilon \leq 0.5$, we have $n^{\dag} < 2$ again due to $f(\gamma_w^{\dag}) > 0.4675$. We next confirm that even for $n^{\dag} < 2$ we still have $n^{\ast} = N$. To this end, we only have to confirm $n \gamma_w$ for $n = 2$ is larger than that for $n = 1$.
When $n = 1$, following \eqref{n_P} we have $f(\gamma_w) = 2 \epsilon^2$. The maximum value of $\gamma_w$ that guarantees $f(\gamma_w) = 2 \epsilon^2$ (i.e., $n = 1$)  is obtained when $\epsilon = 0.5$ since $f(\gamma_w)$ is a monotonically increasing function of $\gamma_w$ as proved by \eqref{f_P_d}. We obtain this maximum value by solving $f(\gamma_w) = 0.5$ as $\gamma_w^{n = 1} < 2.3145$, which leads to $n \gamma_w < 2.3145$ when $n = 1$.
When $n = 2$, following \eqref{n_P} we have $f(\gamma_w) = \epsilon^2$. The minimum value of $\gamma_w$ that guarantees $f(\gamma_w) = \epsilon^2$ (i.e., $n = 2$)  is obtained when $\epsilon = 0.4835$. We obtain this minimum value by solving $f(\gamma_w) = (0.4835)^2$ as $\gamma_w^{n = 2} > 1.16$, which leads to $n \gamma_w >2.32$ when $n = 2$. As such, we have $n \gamma_w < 2.3145$ when $n = 1$ and $n \gamma_w >2.32$ when $n = 2$, which results in $n \gamma_w$ for $n = 2$ is larger than $n \gamma_w$ for $n = 1$. We recall that $n \gamma_w$ monotonically increases with $n$ when $n \geq n^{\dag}$. Therefore, for $0.4835 \leq \epsilon \leq 0.5$ the optimal value of $n$ that maximizes $n \gamma_w$ is $N$.

\textbf{Step 6:}
So far, we have proved $n^{\ast} = N$. Then, substituting $n^{\ast} = N$ into \eqref{n_P}, we obtain the fixed-point equation in \eqref{fix_P}.


\end{document}